\documentclass[twocolumn,preprintnumbers,floatfix,superscriptaddress,nofootinbib]{revtex4}
\pdfoutput=1 
\usepackage{CJKutf8} 
\usepackage{mathtools,slashed,mathrsfs,amsfonts}
\usepackage[caption=false]{subfig}
\usepackage{dcolumn}
\usepackage{multirow}
\usepackage{tabularx}
\usepackage{booktabs}
\usepackage{bm}
\usepackage{breqn}
\usepackage{comment}
\usepackage{setspace}
\usepackage[dvipsnames]{xcolor}
\usepackage[normalem]{ulem} 
\usepackage{enumerate}
\usepackage{siunitx}
\usepackage{multirow}
\usepackage{hyperref}

\newcommand{\eV}{{\, {\rm eV}}}
\newcommand{\keV}{{\, {\rm keV}}}

\definecolor{mypurple}{RGB}{164,64,214}

\newcommand\nn{\nonumber}
\newcommand\eea{\end{eqnarray}}
\newcommand\bea{\begin{eqnarray}}
\newcommand\ees{\end{split}}
\newcommand\bes{\begin{split}}

\def\l{\left(}
\def\r{\right)}

\def\geff{g_{a \gamma \gamma}^{\rm eff}}
\def\g{g_{a \gamma \gamma}}
\def\epeff{\epsilon_{\rm eff}}

\begin{document}
\preprint{LITP-26-10}

\title{A Bandpass Axion \\ {\small Or: How I Learned To Stop Worrying About Stars And Love The Lab}}


\author{Dawid Brzeminski}
\email{dawid@umich.edu}
\affiliation{Leinweber Institute for Theoretical Physics$,$ Department of Physics$,$ \\
University of Michigan$,$ Ann Arbor$,$ MI 48109$,$ USA}
\author{Anson Hook}
\email{hook@umd.edu}
\affiliation{Maryland Center for Fundamental Physics$,$ Department of Physics$,$ \\
University of Maryland$,$ College Park$,$ MD  20742$,$ U.S.A.}


\vspace*{1cm}

\begin{abstract} 

Axion-like particles coupled to photons are one of the most compelling new physics scenarios. We demonstrate that an axion-photon coupling resulting from a non-anomalous PQ symmetry under which light fermions are charged acts as a bandpass filter: both high- and low-energy probes experience a parametrically suppressed coupling while intermediate-energy probes remain unaffected.
An immediate result of this bandpass is that lab-based constraints can naturally be the dominant constraint for almost all values of the axion mass.  High-energy constraints coming from stellar dynamics as well as low-energy constraints coming from photon-axion conversion in galactic/stellar magnetic fields are simultaneously suppressed, while lab-based experiments, such as light-shining-through-a-wall experiments, done at intermediate energies are unsuppressed.

\end{abstract}

\maketitle

\section{Introduction} 

The QCD axion and its cousins, axion-like particles (ALPs), are one of the most well-motivated new physics scenarios.  The QCD axion can solve the strong CP problem~\cite{Peccei:1977ur,Peccei:1977hh,Weinberg:1977ma,Wilczek:1977pj,Zhitnitsky:1980tq,Dine:1981rt,Kim:1979if,Shifman:1979if}, and axions in general can naturally play the role of dark matter~\cite{Preskill:1982cy,Dine:1982ah,Abbott:1982af}. Even independent of these considerations, axions are well-motivated.  Perhaps the best motivation is that we have discovered axion-like particles before, e.g. the pion is an ALP. Additional motivation can be found in the fact that axions are a ubiquitous prediction of extra dimensions and string theory \cite{Svrcek:2006yi,Arvanitaki:2009fg,Demirtas:2018akl}.

There is an active experimental program looking for axions (see {\it e.g.}~\cite{Ringwald:2024uds,AxionLimits,ParticleDataGroup:2024cfk} and references therein).  The non-dark matter searches typically fall into either lab-based experiments or astrophysical constraints, with only a few exceptions.  When looking for light (sub-eV mass) axions, astrophysics appears to hold all of the cards.  Stellar dynamics occur at much higher energies, densities, volume, and magnetic fields than anything available in tabletop experiments; simultaneously, photon-to-axion conversion experiments in outer space enjoy a better vacuum, longer distances and longer coherence lengths.  While lab-based experiments enjoy better precision and control, the many advantages held by astrophysical systems results in much stronger bounds.

From the previous discussion, a natural question arises: ``Are there any axion models for which the dominant constraints come from lab-based experiments?".  
This question was originally answered in Refs.~\cite{Masso:2005ym,Masso:2006gc,Mohapatra:2006pv}, which evaded the CAST and stellar cooling constraints while trying to explain an old PVLAS anomaly~\cite{PVLAS:2005sku}.
Much has changed in the 20 years since these original approaches towards evading astrophysical bounds.  Currently, evading stellar production constraints is not sufficient to render lab-based constraints dominant.  Axions are also constrained by conversion experiments.  Photon-to-axion conversion experiments currently come in two varieties, white dwarf polarization constraints~\cite{Lai:2006af,Gill:2011yp,Dessert:2022yqq,Benabou:2025jcv} and spectral distortions~\cite{Reynolds:2019uqt,Li:2024zst}.  These constraints operate by taking light emitted by some astrophysical body and using the polarization or frequency-dependent conversion rate to place a constraint.  These conversion experiments are a very {\it low}-energy experiment, e.g. the length scale relevant for white dwarfs is $\sim 10^3$ km while the axion-photon conversion lengths scale in the lab is $\sim$ m.  Thus to evade all constraints, the theory of the axion must act like a bandpass, avoiding both very low- and very high-energy probes while allowing intermediate energy probes.

Form factors provide a manner in which to implement such a bandpass.  Form factors typically describe the suppression of a coupling at high energies.
To see how a form factor can suppress both high and low but not intermediate energies, consider an axion coupled to two fermions with the same electric charge but opposite PQ charge as
\bea
m_\psi e^{i a/f} \psi \psi^c +m_\chi e^{-i a/f} \chi \chi^c + h.c. 
\eea
The PQ symmetry does {\it not} have an anomaly with photons so at low energies, $E \ll m_\psi <m_\chi$, there is no photon coupling, $\geff \approx 0$.  The photon coupling coming from $\psi$ is canceled against the photon coupling coming from $\chi$.  At high energies, $E \gg m_\chi > m_\psi$, any effective photon coupling is suppressed by the masses of the fermions, $\geff \propto (m_\chi^2-m_\psi^2)$.  However, at intermediate energies, $m_\psi \lesssim E \lesssim m_\chi$, the photon coupling coming from $\psi$ is suppressed while the photon coupling coming from $\chi$ is still present.  We thus arrive at a scenario where the effective axion-photon-photon coupling is suppressed at low and high energies, but is large in an intermediate regime.  In the rest of the paper, we flesh out the model and its experimental consequences.

\section{Model}

The model we study is a small variation along the lines of what was proposed in the Introduction.  Because we will need $m_\psi \lesssim$ keV to suppress astrophysical conversion constraints, $\psi$ will be constrained by many experimental searches.  In particular, $\psi$ will necessarily be a millicharged particle with a small charge $\epsilon e$ subject to its own very strong astrophysical constraints.  If $\epsilon$ is too small, then $\geff \propto \epsilon^2$ will be too small and lab-based experiments will be unable to reach the sensitivity needed to probe the unsuppressed value of $\geff$.  To evade this constraint, we augment our model with an approach used in Refs.~\cite{Masso:2006gc,Berlin:2020pey}.  The idea is to make $\epsilon$ a function of the effective photon mass $m_\gamma$ so that the millicharge is suppressed in stellar environments.

Other than electromagnetism $U(1)_\gamma$, the model we consider has 2 additional $U(1)$ gauge symmetries $U(1)_1$, $U(1)_2$ and 3 global non-anomalous symmetries $U(1)_{PQ}$, $U(1)_\psi$, $U(1)_\chi$
\begin{center}
\begin{tabular}{c|ccc|ccc}
&$U(1)_\gamma$&$U(1)_1$&$U(1)_2$&$U(1)_{PQ}$&$U(1)_\psi$&$U(1)_\chi$ \\
\hline
$\psi$&$0$&1&-1&-1/2&1&0 \\
$\psi^c$&$0$&-1&1&-1/2&-1&0 \\
$\chi$&$0$&1&-1&1/2&0&1 \\
$\chi^c$&$0$&-1&1&1/2&0&-1 \\
$\Phi$&$0$&0&0&1&0&0 
\end{tabular}
\end{center}
with a Lagrangian
\bea
\mathcal{L} = - \frac{1}{4} F^T \begin{pmatrix}
1 & \epsilon & \epsilon \\
\epsilon & 1 & 0 \\
\epsilon & 0 & 1
\end{pmatrix} F + \frac{1}{2} A^T \begin{pmatrix}
m_\gamma^2 & 0 & 0 \\
0 & m_D^2 & 0 \\
0 & 0 & 0
\end{pmatrix} A + J^T A \nn \\
+ y_\psi \Phi \psi \psi^c + y_\chi \Phi^\dagger \chi \chi^c + \frac{1}{2} m^2 |\Phi|^2 - \frac{\lambda}{4} |\Phi|^4 - V(a) + h.c.\, , \nn
\eea
where $J = (J_{\gamma}, J_1, J_2)^T$, $F = (F_{\gamma}, F_1, F_2)^T$, $A = (A_{\gamma}, A_1, A_2)^T$ and we have included a plasma mass for the photon to indicate how the results change in stellar environments.  Kinetic mixing between $A_1$ and $A_2$ will not change the final results and has been dropped for simplicity.
$\Phi$ obtains a vev and gives an axion as $\langle \Phi \rangle = f_a e^{i a/f_a}$.  Finally, the potential $V(a)$ indicates whatever mechanism by which the axion obtains its potential and, most importantly, its mass.

The kinetic terms for the gauge bosons can be diagonalized to $\mathcal{O}(\epsilon^2)$ with the transformation
\bea
A = \begin{pmatrix}
A_\gamma \\
A_1 \\
A_2
\end{pmatrix} \rightarrow   \begin{pmatrix}
1 & \epsilon \frac{m_D^2}{m_\gamma^2-m_D^2} & 0 \\
-\epsilon \frac{m_\gamma^2}{m_\gamma^2-m_D^2} & 1 & 0 \\
-\epsilon & 0 & 1
\end{pmatrix} A \, .
\eea
This whole process generates the coupling to photons, electrons, $\psi$ and $\chi$ :
\bea
J^T A &\rightarrow& J^T A - e \epsilon_{\rm eff} A_\gamma (J_\psi + J_\chi)  + e \epsilon_{\rm eff} A_1 J_{\gamma} \nn \\
\epsilon_{\rm eff} &=& \epsilon \frac{m_D^2}{m_\gamma^2-m_D^2} \, , \label{Eq: epeff}
\eea
where we have taken the two new $U(1)$s to have the same gauge coupling (for simplicity we choose $e_1=e_2=e$), which amounts to imposing a $\mathbb{Z}_2$ symmetry softly broken by $m_D$.
We see that we have millicharged fermions and a dark photon with photon mass dependent  couplings.  

\begin{figure}[t]
    \centering
    \includegraphics[width=0.49\textwidth]{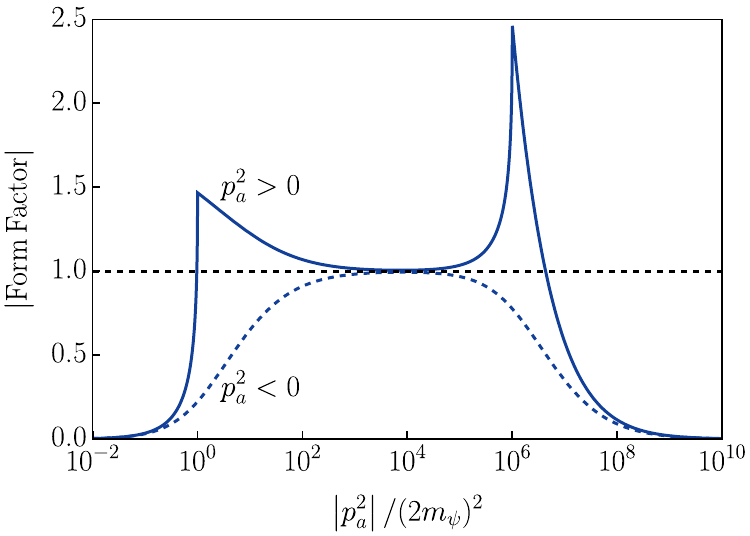}
    \caption{Absolute value of the form factor $F$ with $m_\chi = 10^3 \, m_\psi$ and $p_{\gamma,1}^2=p_{\gamma,2}^2=0$. Since the form factor is symmetric in its three external invariants, the plot also represents any case in which a single $p^2$ dominates over the others.}
    \label{Fig: formfactor}
\end{figure}

What we are interested in is the effective coupling of the axion to photons
\bea
\frac{\geff}{4} a F \tilde F.
\eea
The effective coupling to photons after summing over both fermions is~\cite{Bauer:2021mvw,Ferreira:2022xlw} :
\bea
& &g_{a \gamma \gamma}^{eff} = \g \frac{\epeff^2}{\epsilon^2} F\l\frac{p_{\gamma,1}^2}{4 m_\psi^2},\frac{p_{\gamma,2}^2}{4 m_\psi^2},\frac{p_{a}^2}{4 m_\psi^2},\frac{p_{\gamma,1}^2}{4 m_\chi^2},\frac{p_{\gamma,2}^2}{4 m_\chi^2},\frac{p_{a}^2}{4 m_\chi^2}\r, \nn \\
& &\g = \frac{e^2 \epsilon^2}{4 \pi^2 f_a}, \nn \\
& &F(a,b,c,A,B,C) = f(a,b,c) - f(A,B,C), \\
& &f(a,b,c) =  \int dx  dy dz \frac{-2 \delta(1-x-y-z)}{1 - 4 x z a  - 4  y z b - 4 x y c - i \epsilon} \, . \nn
\eea
This form factor quickly approaches zero at small and large energies, but is non-zero and $\sim 1$ in between $m_\psi$ and $m_\chi$, as shown in Fig.~\ref{Fig: formfactor}.  In practice, we evaluate the form factor using LoopTools~\cite{Hahn:1998yk}.

\section{Constraints}

The bandpass axion has constraints on the axion as well as on the UV completion.  We will consider these separately.  

\subsection{Constraints on the UV completion}

Constraints on the UV completion come in the form of constraints on $\epeff$ from searches for millicharged particles, from dark photons, and from dephasing.  As our main focus is on the bandpass axion, as opposed to the UV completion, we will choose the data points which lie on the edge of current constraints on the UV completion and keep optical LSW experiments unsuppressed.

\paragraph{Dark photon}

\begin{figure}[t]
    \centering
    \includegraphics[width=0.49\textwidth]{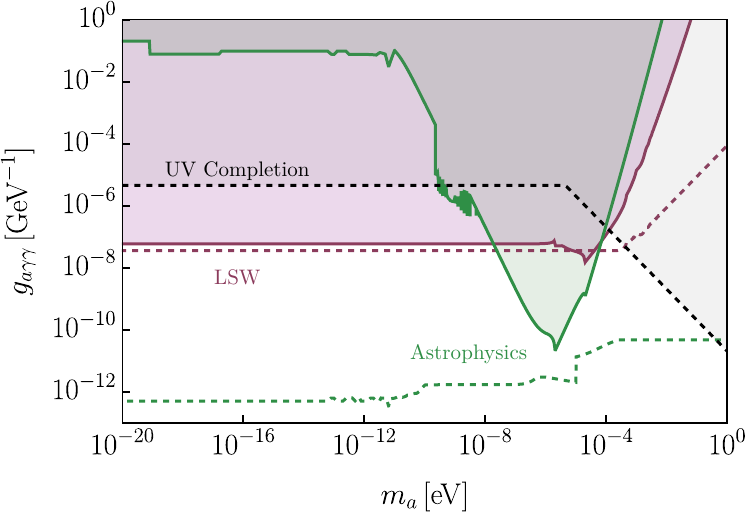}
    \caption{Constraints when $m_\psi = 10^{-6}$ eV, $m_\chi = 10^{-5}$ eV, $\epsilon = 2.5 \times 10^{-9}$ and $m_D = 6 \times 10^{-4} \, {\rm eV}$.  In green we have astrophysical constraints \cite{Dessert:2022yqq,Reynolds:2019uqt,Li:2024zst}, in purple we have the constraints from LSW experiments \cite{Ehret:2010mh,OSQAR:2015qdv,DellaValle:2015xxa} while in dotted black, we have constraints on the UV completion. The green (purple) dashed line is an envelope of all of the astrophysical (LSW) constraints before applying the corrections due to the form factor and modified millicharge~\cite{Reynes:2021bpe,Fermi-LAT:2016nkz,Ning:2024eky,Noordhuis:2022ljw,Ruz:2024gkl,Ayala:2014pea,Dolan:2022kul,Foster:2022fxn,Dessert:2022yqq}. Lab-based experiments are stronger than astrophysical constraints in the mass range $m_a \gtrsim 7 \times 10^{-5}$ eV and $m_a \lesssim 3 \times 10^{-8}$ eV. Constraints requiring stellar production are additionally suppressed by $\epeff$ and do not appear on the plot.
    }
    \label{Fig: constraint}
\end{figure}

In the mass range we are considering, the dominant constraints on dark photons come from stellar cooling~\cite{Vinyoles:2015aba,An:2020bxd,XENON:2021qze}, giving roughly
\bea
\epsilon \, m_D \lesssim 1.5 \times 10^{-12} \, {\rm eV} .
\eea
The constraints on dark photons are stronger than the constraints on the millicharged particles as emission of the longitudinal mode of the dark photon is enhanced.  Using the constraint $f_a \gtrsim m_a$, we can translate this into a curve
\bea
m_a \geff \lesssim \frac{e^2}{4 \pi^2} \left ( \frac{10^{-12} \, {\rm eV} }{m_D} \right )^2 \, .
\eea
This constraint is shown with the diagonal part of the black dashed line in Fig.~\ref{Fig: constraint} and Fig.~\ref{Fig: constraint2} when $m_D = 6 \times  10^{-4}$ eV.
\paragraph{Yukawa couplings}
We require that $y_\psi, y_\chi \lesssim 1$ or equivalently $f_a\gtrsim m_\psi, m_\chi$. In practice, this sets an upper bound on the value of $\g$
\bea
\g \lesssim \frac{e^2\epsilon^2}{4\pi^2 m_\chi} \, ,
\eea
where we take $m_\chi$ to be the larger of the two fermion masses. The constraint appears as the horizontal part of the black dashed line in Fig.~\ref{Fig: constraint} and Fig.~\ref{Fig: constraint2}.  
\paragraph{Millicharged particles}
The leading astrophysical constraint on light millicharged particles comes from stellar cooling and is~\cite{Davidson:2000hf}
\bea
\epeff \lesssim 10^{-14} \quad \to \quad \epsilon \lesssim 1 \l \frac{10^{-4} \eV}{m_D} \r^2 \l \frac{m_\gamma}{1 \keV} \r^2 .
\eea
This constraint is weaker than the constraint coming from dark photons.
The leading lab-based bound on millicharged particles is
\bea
\epsilon <  2 \times 10^{-7} \, ,
\eea 
coming from the PVLAS experiment~\cite{DellaValle:2014xoa}.

\paragraph{Dephasing}

As is familiar from many dark photon models, if $m_D = 0$ then the dark photon decouples entirely. From Eq.~\ref{Eq: epeff} the fermions, and hence the axion, will completely decouple in this limit as well.  
The way that this decoupling manifests itself  is that the light produced by electrons and protons is $A_\gamma + \epsilon A_1$ while the light coupled to $\psi$ and $\chi$ is $\epsilon A_\gamma - A_1 + A_2$, orthogonal to the light produced by electrons.

To see how this decoupling occurs in a light-shining-through-a-wall (LSW) experiment, let us consider the conversion probability of the light produced by electrons into axions in the presence of a magnetic field.  The conversion of an axion back into light will feature the same peculiarities.  A standard calculation~\cite{Raffelt:1987im} gives the probability to be
\bea
p_{\gamma \rightarrow a} = \left|\int_0^L dz \frac{\geff (B_\gamma - B_1/\epeff)}{2} e^{i \l \frac{m_\gamma^2 - m_a^2}{2 \omega} \r z} A_\gamma^0 \nn \right. \\
\left. - \frac{\geff (B_\gamma - B_1/\epeff)}{2 \epeff} e^{i \l \frac{m_D^2 - m_a^2}{2 \omega} \r z} A_1^0 \right|^2 \, , \label{Eq: conversion}
\eea
where $\omega$ is the energy of the incoming light. We have taken there to be two magnetic fields ($B_\gamma$,$B_1 \ne 0$ with $B_2  = 0$), as well as two incoming normalized complex amplitudes ($A_\gamma^0$,$A_1^0 \ne 0$ and $A_2^0 = 0$).  From this, we see two requirements that LSW experiments must satisfy in order to have any observable effects in the $m_D \rightarrow 0$ limit. The first is that the magnetic field produced by electrons (for which $B_1 =  \epeff B_\gamma$) does not allow for any conversion.  Only after a distance $1/m_D$ from the magnet does $B_1$ get exponentially suppressed, allowing axion-photon mixing to occur.  This imposes the requirement $m_D > 1/{\rm m} \sim 10^{-7}$ eV.

A stronger requirement comes from when the incoming light obeys $A_1^0 = \epeff A_\gamma^0$, as light produced by electrons will initially satisfy.  In this case, the leading-order conversion of $A_1$ into axions cancels the conversion of $A_\gamma$ into axions and no axions are produced.  The net result is that light will in fact not shine through a wall.  Axions are produced when this delicate cancellation is disrupted, namely when $A_1$ and $A_\gamma$ are no longer in phase with each other.  This occurs when
\bea
\Delta k \, L \sim \frac{(m_D^2 - m_\gamma^2 )}{2\omega} \, L \gtrsim 1 \, .
\eea
Mathematically, this requirement can be seen explicitly by expanding Eq.~\ref{Eq: conversion} for small $L$.  There one sees that the usual conversion probability $\sim \g^2 B^2 L^2$ is suppressed and becomes $\sim \g^2 B^2 L^2 (\Delta k \, L)^2 $ and only becomes the usual result when $\Delta k \, L \gtrsim 1$.
The leading lab-based experiments use laser light with $\omega \sim$ eV, have $m_\gamma \lesssim 10^{-5} \, \rm eV$, and occur over length scales of $\sim$ 1 meter so we find the requirement
\bea
m_D \gtrsim \sqrt{\frac{2\omega}{\rm eV} \frac{1 \, {\rm m}}{L}} \sim 6 \times 10^{-4} \, {\rm eV} \, .
\eea
For smaller $m_D$, longer-baseline or microwave setups such as CROWS \cite{Betz:2013dza} remain relevant. 

Combining the dark photon bound with the dephasing requirement, we take $\epsilon = 2.5 \times 10^{-9}$ and $m_D = 6\times  10^{-4}$ eV in the figures below. The choice lies near the edge of current dark photon constraints while satisfying the dephasing requirement needed for optical LSW experiments to probe the unsuppressed coupling.
\subsection{Constraints on the bandpass axion}

\begin{figure}[t]
    \centering
    \includegraphics[width=0.49\textwidth]{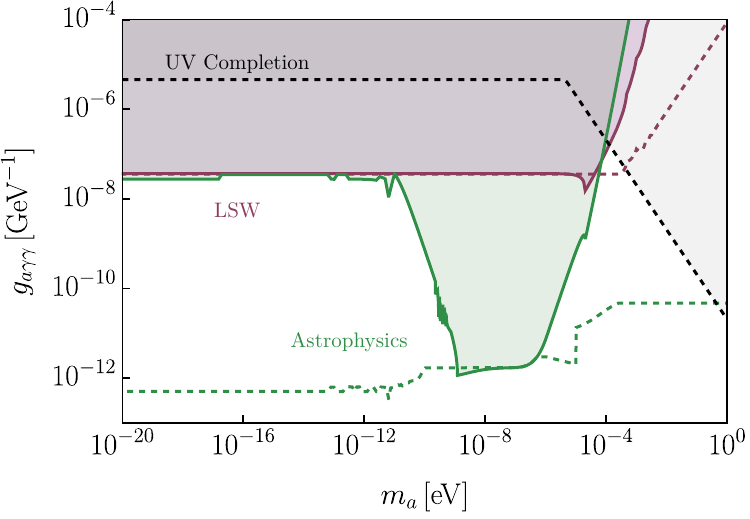}
    \caption{Constraints when $m_\psi = 6 \times 10^{-10}$ eV and $m_\chi = 10^{-5}$ eV, $\epsilon = 2.5 \times 10^{-9}$ and $m_D = 6 \times 10^{-4} \, {\rm eV}$.  The colors are the same as in Fig.~\ref{Fig: constraint}. For this choice of fermion mass, the suppression of astrophysical limits makes the astrophysical and lab-based constraints comparable in the low-mass limit. Constraints requiring stellar production are removed from the plot by the additional environmental suppression of $\epeff$.}
    \label{Fig: constraint2}
\end{figure}

The couplings of bandpass axions can be suppressed by two effects, the form factor $F$ and the the effective millicharge $\epeff$.  

\paragraph{Stellar production}

For stellar production of any sort, we calculate $\epeff$ using $m_D = 6 \times  10^{-4}$ eV and $m_\gamma$ of the stellar environment. For these bounds, the dominant suppression in our benchmarks is the environmental millicharge suppression. This suppresses cooling bounds by a factor $(\epsilon_{\rm eff}/\epsilon)^2 \sim 10^{-25}$, and CAST by a factor of $(\epsilon_{\rm eff}/\epsilon) \sim 10^{-12} $. Given that the form factor $\lesssim 1$, constraints that require stellar production are so suppressed that they do not appear in the plots in Fig.~\ref{Fig: constraint} and Fig.~\ref{Fig: constraint2}.

\paragraph{Conversion}
Most searches for photon-to-axion conversion occur in regions where $m_\gamma \ll m_D$ so there is no $\epeff$ suppression. However, the form factor suppression remains. In this case, we have
\bea
p_{\gamma,1}^2 = m_\gamma^2 \quad p_{\gamma,2}^2 = - {\rm max} \, \left ( \frac{1}{L} , \frac{m_\gamma^2- m_a^2}{2 \omega} \right )^2 \quad p_a^2 = m_a^2 . \nn
\eea
As in stellar production, photon-to-axion conversion has an on-shell photon and axion.  Here, the magnetic field supplies momentum, whose value is either the minimum required by conservation of energy, $\sim (m_\gamma^2- m_a^2)/2\omega$, or a non-minimal amount $k \sim 1/L$.  $L$ is the shortest length scale in the problem.  e.g. the length scale over which the B field or $m_\gamma$ changes.  For most situations, one can just use the length scale of conversion as it is typically, but not always, a proxy for the shortest length scale.  

The value of $m_\gamma$ is what separates low-energy astrophysics from lab-based experiments.  At ``high" axion masses, $m_a \gtrsim m_\psi, m_\gamma$, the dominant momentum flowing through the vertex is $m_a^2$.  In this case, lab and astro constraints experience the same form factor.  Meanwhile, at low axion masses, $m_a \lesssim m_\psi,m_\gamma$, the dominant momentum flowing through the form factor is $p_{\gamma,1}^2 = m_\gamma^2$ giving a $m_\gamma^2/m_\psi^2$ suppression.  The vacuum of outer space is good and $m_\gamma \sim 10^{-11}$ eV, while typical lab-based experiments have $m_\gamma \sim 10^{-5,-6}$ eV.

There are three main astrophysical conversion experiments that place relevant constraints.  In all of these scenarios the largest energy scale in the problem is $m_\gamma$.  Firstly, there are magnetic white dwarfs constraints.   Ref.~\cite{Dessert:2022yqq} argues that $m_\gamma \lesssim 10^{-11}$ eV.  We take $m_\gamma$ to saturate this inequality.  The second conversion constraint comes from X-ray transparency of the intergalactic medium~\cite{Marsh:2017yvc,Reynolds:2019uqt}.  For these experiments $m_\gamma$ is spatially dependent. Since the bounds depend on $m_\gamma^2-m_a^2$, one can estimate the value of $m_\gamma$ in the region dominating the bound by determining where the bound first becomes sensitive to $m_a$. Hence we take $m_\gamma = 10^{-11}$ eV.  The last relevant conversion constraint comes from spectral distortions of a nearby blazar~\cite{Li:2024zst} for which $m_\gamma = 10^{-8}$ eV.

Finally, there are lab-based constraints coming from ALPS, PVLAS, and OSQAR.  In these experiments, the effective photon mass is induced by the residual gas in the imperfect vacuum. We estimate the corresponding index of refraction $n$ from the reported upper limits on the residual gas pressure $P$,  using $n-1 \approx 3 \times 10^{-4} \, (P/\rm bar)$ for a dilute atmospheric gas. The resulting contribution to the photon dispersion can be parametrized as  $m_{\gamma}^2 = - 2(n-1)\omega^2$. We take $\vert m_{\gamma,\, \rm ALPS}\vert = 6 \times 10^{-6} \, \rm eV$, $\vert m_{\gamma,\, \rm PVLAS} \vert = 6 \times 10^{-7} \, \rm eV$, and $\vert m_{\gamma,\, \rm OSQAR} \vert = 6 \times 10^{-6} \, \rm eV$, estimated from upper limits on the vacuum pressures reported in Refs.~\cite{ALPS:2009des,DellaValle:2015xxa,OSQAR:2015qdv}.  In Fig.~\ref{Fig: constraint}, we see that lab-based constraints can be the leading source of constraints for a large region of parameter space. Even if the residual gas contribution is neglected, the finite size of the magnetic field region probes $p_{\gamma, 2}^2\sim - 1/L^2$ in the low axion mass limit yielding similar plots. For 1 m scale experiments like PVLAS, the form factor is $\mathcal{O}(1)$ when $m_\psi \lesssim 2\times 10^{-7} \, {\rm eV} \lesssim m_\chi$.

The effect of differing $m_\psi$ can be seen by comparing Fig.~\ref{Fig: constraint} and Fig.~\ref{Fig: constraint2}.  For larger $m_\psi$, astrophysical constraints are suppressed by a stronger form factor.  In both cases, lab-based constraints are unsuppressed at low axion masses. In particular, Fig.~\ref{Fig: constraint2} demonstrates that a minimal value of $m_\psi \gtrsim 6 \times 10^{-10} \, \rm eV$ is required for astrophysical constraints to be subleading to lab-based limits.

\section{Conclusion}

In this article we explored the bandpass axion; namely, an axion whose PQ fermions are light.  The resulting form factor disrupts a delicate symmetry-enforced cancellation, resulting in the bandpass axion, which suppresses the effective coupling seen by high-energy production in stellar environments as well as the low-energy conversion constraints in astrophysical magnetic fields, but could still be discovered in lab-based experiments conducted at intermediate energy and length scales.  This demonstrates that lab-based axion experiments place the leading bounds over a wide range of masses on some axion models.

Interestingly enough, there are two mass regions where lab-based constraints beat stellar constraints, which for the data point in Fig.~\ref{Fig: constraint} are $m_a \gtrsim 7 \times 10^{-5}$ eV and $m_a \lesssim 3 \times 10^{-8}$ eV.  The region around $m_a \gtrsim 7 \times 10^{-5}$ eV is sensitive to the UV completion and constraints on it, e.g. the axion may decay before converting back into a photon.  It would be interesting to explore other models in which this region is more robustly allowed.  It would also be interesting to explore if there are other models in which lab-based constraints are the leading constraint over all of parameter space.

Finally, it would be interesting to explore in more detail axion constraints coming from neutron stars~\cite{Noordhuis:2022ljw}.  Neutron stars are unique in that the length scales involved may approach lab-based length scales, giving them the sole opportunity of being an astrophysical constraint probing lab-based momentum transfers.

\section*{Acknowledgments}

We thank Zackaria Chacko and Abhishek Bannerjee for inspiring discussions and Prateek Agrawal, Gustavo Marques-Tavares, and Aaron Pierce for comments on the draft and title.  AH is supported by NSF grant PHY-2514660 and the Maryland Center for Fundamental Physics. DB is supported by the Department of Energy under grant number DE-SC0007859.

\bibliography{biblio}{}

\bibliographystyle{JHEP}

\end{document}